\documentstyle[12pt,aasms4]{article}

\lefthead{Crenshaw, Maran, \& Mushotzky}
\righthead{Intrinsic C IV Absorption in NGC 3516}

\begin{document}

\title{Resolving the Intrinsic C~IV Absorption in the Seyfert 1 Galaxy 
NGC~3516\altaffilmark{1}}

\author{D. Michael Crenshaw\altaffilmark{2}, Stephen P. Maran\altaffilmark{3,4}, 
and Richard F. Mushotzky\altaffilmark{5}} 

\altaffiltext{1}{Based on observations with the NASA/ESA Hubble Space 
Telescope, obtained at the Space Telescope Science Institute, which is 
operated by the Association of Universities for Research in Astronomy, 
Inc., under NASA contract NAS5-26555.}

\altaffiltext{2}{Catholic University of America,
Laboratory for Astronomy and Solar Physics,
Code 681, NASA/Goddard Space Flight Center,
Greenbelt, MD  20771.}

\altaffiltext{3}{Goddard High-Resolution Spectrograph (GHRS) Investigation 
Definition Team.}

\altaffiltext{4}{Space Sciences Directorate, Code 600,
NASA/Goddard Space Flight Center, Greenbelt, MD  20771.}

\altaffiltext{5}{Laboratory for High Energy Astrophysics,
Code 666, NASA/Goddard Space Flight Center, Greenbelt, MD  20771.}

\begin{abstract}

We observed the Seyfert 1 galaxy NGC~3516 with the Goddard High Resolution 
Spectrograph on the Hubble Space Telescope, and obtained UV spectra at a 
resolution of $\lambda$/$\Delta\lambda$ $\approx$ 20,000 in the redshifted C~IV 
$\lambda$1549 region. The intrinsic C~IV absorption in the core of the broad 
emission line is resolved for the first time into four distinct kinematic 
components, which are all blue-shifted relative to the systemic radial velocity 
of the host galaxy. Two components are narrow ($\sim$20 and $\sim$30 km s$^{-1}$ 
FWHM) and at small radial velocities ($-$90 and $-$30 km s$^{-1}$ respectively), 
and could arise from the interstellar medium or halo of the host galaxy. The two 
broad components are centered at radial velocities of $-$380 and $-$150 km 
s$^{-1}$, have widths of 130 and 210 km s$^{-1}$ FWHM respectively, and most 
likely arise in outflowing gas near the active nucleus. At the times of 
observation, 1995 April 24 and 1995 October 22, there was no evidence for the 
variable absorption component at a higher outflow velocity that disappeared some 
time after 1989 October (Koratkar et al. 1996). 

The cores of the broad C~IV absorption components are very close to zero 
intensity, indicating that the absorption regions are extended enough to 
completely occult the broad C~IV emitting region (which is $\sim$9 light days 
in extent). The total column density of the C~IV absorption is substantially 
larger than that measured by Kriss et al. (1996a) from contemporaneous HUT 
spectra, but we agree with their conclusion that the UV absorption is too strong 
to arise primarily from the X-ray warm absorber region. The
GHRS observations separated by six months reveal no appreciable changes in the 
equivalent widths or radial velocities of any of the absorption components, 
although a change in the column density of either broad component at a level 
$\leq$~25\% cannot be ruled out, because the lines are highly saturated. At 
present, we know of two Seyfert galaxies, NGC 3516 and NGC 4151 (Weymann et al. 
1997), with complex multiple absorption zones that can remain stable in column 
density and velocity field over time scales of months to years.

\end{abstract}

\keywords{galaxies: individual (NGC 3516) -- galaxies: Seyfert}

\section{Introduction}

NGC 3516 is one of the few Seyfert 1 galaxies with intrinsic absorption lines 
that were strong enough to be detected by the International Ultraviolet 
Explorer (IUE). The UV absorption is characterized by high-ionization resonance 
lines (C~IV $\lambda\lambda$1548.2, 1550.8; N~V $\lambda$1238.8, 1242.8; and Si 
IV $\lambda\lambda$ 1393.8, 1402.8) that are blueshifted with respect to the 
host galaxy. The absorption lines were originally studied by Ulrich and Boisson 
(1983) and were recognized as being intrinsic to the nucleus from their large 
widths (extending from zero to $-$3000 km s$^{-1}$ relative to the redshift of 
the host galaxy), equivalent widths (up to $\sim$4 \AA\ for C~IV), and 
variability (on time scales of months). Subsequent IUE monitoring found evidence 
for absorption-line variations on time scales as small as weeks (Voit et al. 
1987; Walter et al. 1990; Kolman et al. 1993).

Walter et al. (1990) characterized the C~IV absorption in NGC 3516 as a blend of 
a narrow stable component near the center of the broad C~IV emission profile, 
and a variable and broad blueshifted component. Koratkar et al. (1996) found 
that the variable component was present in IUE spectra since the first IUE 
observation in 1978, but disappeared between 1989 October and 1993 
February, when there were no IUE observations. More recent spectra obtained with 
the Hopkins Ultraviolet Telescope (HUT) show that the variable blueshifted 
component was not present in 1995 March (Kriss et al. 1996a).

NGC 3516 also exhibits strong and variable X-ray absorption in the form of a 
``warm absorber'' (Kolman et al. 1993; Nandra \& Pounds 1994; Kriss et al. 
1996b; Mathur et al. 1997).  The warm absorber is highly ionized gas 
characterized by high ionization parameters (U $\approx$ 1 -- 10) and 
correspondingly high temperatures (T $\approx$ 10$^{5}$K), and is most easily 
identified from the presence of O~VII and O~VIII absorption edges (see Reynolds 
\& Fabian 1995, and references therein). UV absorbers and X-ray warm absorbers 
are thought to be related, because they both occur in the same objects (Crenshaw 
1997), but the exact nature of the relationship is unclear. In some AGN, they 
could both arise from gas characterized by a single ionization parameter (Mathur 
1994; Mathur et al. 1996). However multi-component models that span a wide range 
in ionization parameter are needed to explain the column densities of the UV and 
X-ray absorption features in NGC 3516 (Kriss et al. 1996a) and NGC 4151 (Kriss 
et al. 1995; Weymann et al. 1997).

Since none of the previous UV observations of NGC~3516 have resolved 
the velocity structure in the C~IV doublet, we decided to observe this 
region at a spectral resolution at least ten times that of any prior 
observation. The observations are part of an ongoing study to understand the 
nature of the intrinsic absorption in low-redshift Seyfert galaxies; results on 
the variable C~IV absorption in NGC~3783 have already been published (Maran et 
al. 1996). For this investigation, we obtained observations of NGC~3516 with the 
Goddard High Resolution Spectrograph (GHRS) on two occasions in 1995 separated 
by six months. 

\section{Observations and Direct Measurements}

We observed NGC~3516 with the GHRS side 2 detector and G160M grating on 1995 
April 24 and 1995 October 22. (The GHRS was removed from the HST during the 
Second Servicing Mission in 1997 February.) The observations were made in 
conjunction with the corrective optics package COSTAR through the Large Science 
Aperture (1$''\!.$74 x 1$''\!.$74), providing a spectral resolution of 
$\lambda$/$\Delta\lambda$ $\approx$ 20,000 ($\lambda$ is the observed wavelength 
and $\Delta\lambda$ is the FWHM of an instrumental profile). For each 
observation, we used two grating wheel settings to cover the wavelength regions 
1528.5 -- 1564.6 \AA\ and 1561.3 -- 1597.3 \AA. Exposure times on the target 
were 121.6 min per setting for the 1995 April spectra and 101.4 min per setting 
for the 1995 October spectra.

We reduced the spectra using the IDL procedures written for the GHRS Instrument 
Definition Team (Blackwell et al. 1993). For each readout, an average background 
across the diode array was determined and subtracted from the gross spectrum, 
since the background level (due primarily to Cerenkov radiation) is constant to 
within a few percent as a function of diode, but generally variable by a factor 
of two as the geomagnetic latitude of HST changes over time (Ebbets 1992). In 
the regions of overlap, the average fluxes from the two wavelength settings 
agree to within 4\%, so no scaling was done between them. The signal-to-noise 
ratio (per  half resolution element) ranges from 8 to 15 for the combined 1995 
April spectrum and 6 to 13 (wings to peak) for the combined 1995 October 
spectrum. 

Figure 1 shows the two GHRS spectra (smoothed by a five-point boxcar for display 
purposes), which encompass most of the broad C~IV emission profile. All of 
the significant absorption features in Figure 1 are due to the resolved C~IV 
$\lambda\lambda$1548.2, 1550.8 doublet at different radial velocities, including 
the strong Galactic lines in the blue wing of the emission profile. The 
intrinsic C~IV absorption is located in the core of the emission profile,
and is clearly separated into a number of distinct components. The variable 
highly blue-shifted component that disappeared after 1989 is still not present 
in these spectra.

In order to determine the strength of the absorption components, the shape of 
the underlying emission must be characterized.  The 1995 April spectrum is 
at a higher flux level, primarily as a result of a continuum level that is 
$\sim$3 x 10$^{-14}$ erg s$^{-1}$ cm$^{-2}$ \AA$^{-1}$ higher (as estimated from 
the far red wing). The broad emission profiles are asymmetric, with more 
emission in the blue wing than the red wing. In addition, the 1995 April profile 
has relatively more emission just blueward of the core than the 1995 October 
profile. To estimate the underlying broad emission in the core of each profile, 
we fit a cubic spline to regions on either side of the core; the fits are shown 
as dotted lines in Figure 1. Note that the flux may be underestimated if there 
is a significant narrow-line contribution to the C~IV emission.

We normalized the absorption lines in the core of the emission profile by 
dividing by the cubic spline fits to the emission core (plus underlying 
continuum); the resulting (unsmoothed) absorption profiles are shown in Figure 
2. The intrinsic C~IV absorption can be separated into four distinct kinematic 
components. The radial velocities of the kinematic components are such that 
there is essentially no overlap in wavelength of the C~IV doublet lines. The 
components that we have labeled 1 and 2 are broad and very deep, approaching 
zero intensity. Components 3 and 4 are narrow, but are clearly seen in each 
member of the doublet on both occasions. The absorption profiles obtained on the 
two dates are visually quite similar.

We made direct measurements of the centroid, equivalent width (EW), and 
full-width at half-maximum (FWHM) of each absorption component -- these 
values are listed in Table 1. Each line is identified as a member of the C~IV 
doublet, and its heliocentric radial velocity (cz) from the line's 
centroid is given. There is no evidence that components 3 and 4 have broad 
wings, so we determined a baseline for measurements of these components from 
adjacent regions in the red wing of component 2. Errors in EW are the sums in 
quadrature of the uncertainties (one standard deviation) from photon noise and 
the estimated errors from different reasonable placements of the fit to the core 
of the emission line. Errors for the other measurements were estimated in the 
same way, and yield values around $\pm$10 km s$^{-1}$ for both FWHM and radial 
velocity. Given the errors, there is no evidence for changes in position, width, 
or equivalent width of any of the components over a six-month interval.

\section{Optical Depths and Covering Factors}

Here we present the details of determining the column densities of the 
intrinsic absorption components from their optical depths and covering factors. 
The instrumental profile has a FWHM of 15 km s$^{-1}$, so each of the C~IV 
absorption components in Table 1 is resolved. Hence we can determine the column 
density of each intrinsic component by integrating its optical depth as a 
function of radial velocity across the profile (Savage \& Sembach 1991). The 
radial velocities are determined with respect to the systemic radial velocity of 
the host galaxy, which we take to be cz $=$ 2634 km s$^{-1}$ from long-slit 
spectra of the narrow H$\alpha$ emission (Keel 1996). This value is in 
good agreement with cz $=$ 2649 km s$^{-1}$ from the stellar absorption 
lines (Vrtilek \& Carleton 1985). The optical depth is just 
\begin{equation}
\tau(v_{r})~=~ln~[F_{0}(v_{r})/F(v_{r})],
\end{equation}
where $F_{0}(v_{r})/F(v_{r})$ is the ratio of continuum to observed flux, or the 
inverse of the normalized flux in Figure 2. To avoid the problem of occasional 
negative values for the normalized fluxes in Figure 2 due to the noise level, we 
binned the fluxes in the cores of components 1 and 2 to 0.14~\AA\ intervals 
($\sim$2 resolution elements), and kept the original bin widths (0.0175 \AA) for 
the remaining portions of the profiles.

Figure 3 shows plots of the observed optical depths as a function of radial 
velocity. Each of the four kinematic components can be easily distinguished in 
these plots. It is obvious that the broad components are {\it not} optically
thin, with $\tau(v_{r})$ $\approx$ 4 -- 6 in the cores, so assuming that the 
lines are unsaturated would severely underestimate the column densities. Errors 
in the optical depths were determined from propagation of the original errors 
(section 2), and can be quite large in the cores despite the large bins. Given 
the errors, the apparent differences in structure in the cores of the lines are 
not significant. 
 
The ratio of C~IV $\lambda$1548.2 to C~IV $\lambda$1550.8 optical depth should 
be 2, since the optical depths are proportional to f$\lambda$ (where f is the 
oscillator strength). However, Figure 3 shows that the ratios are $\sim$1.6 and 
$\sim$1.2 in the cores of components 1 and 2 respectively. This can be explained 
by an additional unabsorbed contribution to the flux, which alters the observed 
optical depth ratio. The two most likely explanations for an unabsorbed 
contribution are: 1) the background emission is partially covered, or 2) there 
is an instrumental contribution (i.e., grating scattered light). In either 
case, the unabsorbed flux can be determined by forcing the ratio of optical 
depths for the C~IV doublet to be 2, and using the formulae given by Hamann et 
al. (1997) to calculate the covering factors in the line of sight (C$_{f}$) and 
corrected optical depths as a function of radial velocity.

We calculated the covering factor for each component and,
given the signal-to-noise, found no evidence for variation of covering factor 
with radial velocity. We therefore determined average covering 
factors (and their dispersions) for each component; the broad component values 
are listed in Table 2 and the narrow component values are discussed below. It is 
obvious from inspection of Figure 2 that the covering factors for the broad 
components are close to one, but Table 2 shows that they are not identical 
to one. Since the broad components approach zero intensity in their cores, even 
a small contribution from unabsorbed light can alter their observed optical 
depths significantly. To demonstrate this point, we used the average covering 
factor for each broad component to calculate the corrected optical depths as a 
function of radial velocity. Then we integrated both observed and corrected 
optical depths over radial velocity for each 
component, and computed the ratio:
\begin{equation}
R_{\tau} = \int \tau_{corrected}(v_{r}) dv_{r} / \int \tau_{observed}(v_{r}) 
dv_{r}.
\end{equation}
This is the factor by which the column density increases when a correction is 
made for a partial covering factor. From Table 2, one can see that this 
correction can be quite significant for these data, particularly for the 
$\lambda$1548.2 component. We conclude that correction for unabsorbed light is 
important for reconciling the doublet optical depths and determining the correct 
column densities of the broad absorption components in NGC 3516. 

The narrow components, which we separated from broad component 2, are noisy and 
yield only approximate values: C$_{f}$ $\approx$ 1.0 $\pm$ 0.3. Since 
the narrow components are shallow, R$_{\tau}$ is roughly proportional to 
1/C$_{f}$. Thus, we adopt a value of C$_{f}$ $=$ 1 for the narrow components, 
with the understanding that the error in optical depth due to uncertain covering 
factor is $\sim$30\%.

Since the covering factors for the components are all suspiciously close to one, 
and scattered light is a basic property of gratings, we conclude that the 
excess light is due to instrumental scattering, which is at the level of 2.5 
$\pm$ 1.5\% (the average value of 1~$-$~C$_{f}$). This value is in agreement 
with the value of $\sim$1.8\% determined by Ebbets (1992), which was calculated 
from GHRS G160M spectra of interstellar absorption lines at 1670 \AA. Thus, we 
believe that the true covering factor for each of the intrinsic C~IV components 
is essentially one (with much higher certainty for the broad components), and we 
have used the ``effective'' covering factors in Table 2 to correct the 
optical depths of the broad components for grating scattered light.  

We obtained the C~IV column densities by integrating the 
corrected optical depths across each component (Savage \& Sembach 1991):
\begin{equation}
N(CIV) = \frac{m_{e}c}{\pi{e^2}f\lambda}~\int \tau(v_{r}) dv_{r},
\end{equation}
and we obtained the errors in column densities from the combination of errors in 
observed optical depths and covering factors. Our final values of the radial 
velocities (relative to the systemic velocity) and widths (FWHM) are averages 
based on the doublet from each kinematic component.

\section{Properties of the Intrinsic Absorption}

For the first time, we can determine properties such as covering factor, column 
density, radial velocity, and variability of the individual kinematic 
components of the C~IV absorption in NGC 3516. We have already shown that 
the covering factor for each absorption component is essentially one and we know 
from IUE monitoring of the continuum and emission lines in NGC 3516 that the 
size (i.e. radius) of the C~IV emitting region is $\sim$4.5 light days (Koratkar 
et al. 1996). Thus, the extent of the individual absorption regions in the plane 
of the sky must be $\geq$ 9 light days. In addition, the UV absorbing regions 
must lie completely outside of the broad C~IV emitting region, at distances 
$\geq$4.5 light days from the continuum source. 

Table 3 provides a summary of the other important properties of the C~IV 
absorption in NGC 3516. Components 3 and 4 have relatively small column 
densities, small radial velocities, and narrow widths. These components could 
easily arise from the interstellar medium or halo of the host galaxy; their 
properties are less extreme than those of the C~IV absorption lines in our 
Galaxy (Table 1). Components 1 and 2 are much stronger and broader, and at 
higher radial velocities relative to the nucleus. These properties are all 
suggestive of outflowing gas close to the nucleus.

Comparing the two GHRS observations separated by six months, there is no 
evidence in Table 3 for variations in any of the properties of any component 
(given our estimates of the errors). The radial velocity coverage of the 
absorption components is also the same for the two observations (see Figure 2), 
which indicates a very stable velocity field over this six-month period. There 
is also no evidence for changes in the column density of any component, although 
the errors indicate that a variation of $\sim$25\% or less cannot be 
ruled out for any component.

Kriss et al. (1996a) obtained HUT spectra of NGC 3516 on 1995 March 11 and 13 
(as well as near-simultaneous X-ray spectra with ASCA). The C~IV doublet is 
resolved in the HUT spectra ($\lambda$/$\Delta\lambda$ 
$\approx$ 500), but the individual kinematic components are not. 
These authors determine a total C~IV column density of 4.7 x 10$^{14}$ 
cm$^{-2}$, whereas if we sum our C~IV column densities over all of the 
components in the GHRS spectra, we obtain values that are $\sim$5 times higher. 
Since the HUT observations were obtained only one month prior to our first 
observation, and there is no evidence for variations six months later, it is 
unlikely that our higher values are due to a real variation in the column 
density. It is more likely that they are due to our higher spectral resolution, 
which allows us to determine the shape of the underlying emission profile more 
accurately, and more importantly, detect the large optical depths in the broad 
components. (To support this claim, we note that the column densities in Table 3 
are $\sim$3 times higher than those that would be determined from the GHRS 
spectra assuming unsaturated lines.) In general, the GHRS data show the 
importance of high spectral resolution for estimating the underlying emission, 
resolving the velocity structure, and directly determining the column densities 
of the intrinsic absorption components from their optical depths. 

It is tempting to associate the UV absorbers with the X-ray warm absorber in NGC 
3516. For some active galaxies, the UV and X-ray column densities have been 
successfully matched with photoionization models characterized by a single 
photoionization parameter, assuming a ``typical'' XUV spectrum and standard 
cosmic abundances (Mathur et al. 1994; Mathur, Wilkes, \& Elvis 1995). 
However, the {\it total} column densities of the UV lines in NGC 3516 and NGC 
4151 are much too high to arise entirely in the highly-ionized warm absorber 
gas, based on the analyses of Kriss et al. (1995, 1996a). With the ability to 
resolve the C~IV absorption, we can ask if any of the individual kinematic 
components are likely to arise from the warm absorber. Components 3 and 4 have 
C~IV column densities that are a factor of ten lower than the value of $\sim$5 x 
10$^{14}$ predicted by the warm absorber model of Kriss et al. (1996a), and as 
we pointed out earlier, they can be explained as absorption from the host 
galaxy. Components 1 and 2 have a combined C~IV column density that is $\sim$ 5 
times higher than their predicted value. Component 1 alone has a 
column density that is close to their predicted value, but there is no other 
evidence that it arises from the warm absorber.
We also note the X-ray warm absorber appears to have varied in the year prior to 
these observations, long after the variable UV component had vanished. 
Comparsion of two ASCA observations on 1994 April 2 (George et al. 1997) and 
1995 March 11 -- 12 (Kriss et al. 1996b) indicate that the optical depths of the 
O VII and O VIII edges were higher by a factor of two on the later date (George 
1997). Thus, there is no clear connection between the UV and X-ray absorbers in 
this Seyfert.

\section{Conclusions}

The intrinsic absorption in NGC 3516 most closely resembles that in the Seyfert 
1 galaxy NGC 4151, in that its UV absorption lines are characterized by very 
high column densities and a number of kinematic components, although the 
absorption in NGC 3516 does not extend to very low ionization lines such as 
Mg~II (Koratkar et al. 1996). Weymann et al. (1997) monitored the intrinsic 
C~IV absorption in NGC 4151 with the GHRS, and were able to place important 
constraints on the radial acceleration of the gas. They find at least 8 distinct 
components at radial velocities from 0 to $\sim$1600 km s $^{-1}$, spanning a 
range of widths and depths. The absorption components in NGC 4151 show little or 
no variation in equivalent width or radial velocity over a four-year period, 
like the observed components in NGC 3516 over a six-month period.  
Other Seyfert galaxies with known intrinsic absorption have lower C~IV 
column densities, and those that have been monitored show evidence for 
equivalent width variations (Maran et al. 1996; Crenshaw 1997). At earlier 
epochs, NGC 3516 itself showed this behavior in the variable blueshifted 
component (Koratkar et al. 1996). 

Intrinsic UV absorbers in Seyfert galaxies share a number of common properties, 
including outflow relative to the host galaxy, high ionization (C~IV, N~V), and 
broad profiles (100 km s$^{-1}$ or more) that separate into more than one 
component at high resolution (Crenshaw 1997). However, there is also a wide 
range in some properties, including column density, variability, and extension 
to low ionization. Some properties, such as covering factors and the fraction of 
active galaxies with intrinsic absorption, are not well known. High-resolution 
UV spectra of a number of absorption lines spanning a wide range in ionization 
level, along with concurrent X-ray spectra, are needed to determine 
the relationship between the multiple ionization and kinematic absorption 
components in NGC 3516, as well as those in other Seyfert galaxies.

\acknowledgments

\clearpage
 
\clearpage
\begin{deluxetable}{ccclcc}
\tablecolumns{6}
\footnotesize
\tablecaption{C~IV absorption components in NGC 3516 \label{tbl-1}}
\tablewidth{0pt}
\tablehead{
\colhead{$\lambda$$_{obs}$} & \colhead{EW} & \colhead{FWHM}
& \colhead{Identification} & \colhead{cz\tablenotemark{a}}
& \colhead{Component}\\
\colhead{(\AA)} & \colhead{(\AA)} & \colhead{(km s$^{-1}$)}
& \colhead{km s$^{-1}$)} &\colhead{(km s$^{-1}$)} & \colhead{} 
}

\startdata
\multicolumn{6}{c}{1995 April 24}\nl
\tableline
1547.85  &0.29$\pm$0.04  &~75  &C IV $\lambda$1548.2  &~$-$68 & Gal.\nl
1550.49  &0.19$\pm$0.03  &~85  &C IV $\lambda$1550.8  &~$-$54 & Gal.\nl
1559.87  &0.75$\pm$0.04  &138  &C IV $\lambda$1548.2  &2259 & 1\nl
1561.04  &1.11$\pm$0.04  &220  &C IV $\lambda$1548.2  &2486 & 2\nl
1561.34  &0.04$\pm$0.02  &~17  &C IV $\lambda$1548.2  &2544 & 3\nl
1561.63  &0.08$\pm$0.02  &~29  &C IV $\lambda$1548.2  &2600 & 4\nl
1562.44  &0.59$\pm$0.02  &121  &C IV $\lambda$1550.8  &2255 & 1\nl
1563.63  &0.99$\pm$0.02  &207  &C IV $\lambda$1550.8  &2485 & 2\nl
1563.96  &0.04$\pm$0.02  &~17  &C IV $\lambda$1550.8  &2549 & 3\nl
1564.23  &0.07$\pm$0.02  &~25  &C IV $\lambda$1550.8  &2601 & 4\nl
\tableline
\multicolumn{6}{c}{1995 October 22}\nl
\tableline
1547.94  &0.26$\pm$0.06  &~62  &C IV $\lambda$1548.2  &~$-$51 & Gal.\nl
1550.52  &0.16$\pm$0.05  &~81  &C IV $\lambda$1550.8  &~$-$49 & Gal.\nl
1559.87  &0.68$\pm$0.03  &127  &C IV $\lambda$1548.2  &2259 & 1\nl
1561.06  &1.16$\pm$0.05  &207  &C IV $\lambda$1548.2  &2490 & 2\nl
1561.35  &0.06$\pm$0.04  &~23  &C IV $\lambda$1548.2  &2546 & 3\nl
1561.65  &0.09$\pm$0.03  &~34  &C IV $\lambda$1548.2  &2604 & 4\nl
1562.44  &0.56$\pm$0.03  &119  &C IV $\lambda$1550.8  &2255 & 1\nl
1563.62  &1.03$\pm$0.04  &190  &C IV $\lambda$1550.8  &2483 & 2\nl
1563.95  &0.03$\pm$0.02  &~23  &C IV $\lambda$1550.8  &2547 & 3\nl
1564.26  &0.06$\pm$0.03  &~36  &C IV $\lambda$1550.8  &2607 & 4\nl
\tablenotetext{a}{Adopted recession velocity for the host galaxy is 2634 km 
s$^{-1}$ (Keel 1996).}
\enddata
\end{deluxetable}

\clearpage
\begin{deluxetable}{cccc}
\tablecolumns{3}
\footnotesize
\tablecaption{Covering factors for the broad components \label{tbl-2}}
\tablewidth{0pt}
\tablehead{
\colhead{Component} & \colhead{C$_{f}$} & \colhead{R$_{\tau}$($\lambda$1548.2)}
& \colhead{R$_{\tau}$($\lambda$1550.8)}
}

\startdata
\multicolumn{4}{c}{1995 April 24}\nl
\tableline
1 & 0.993 ($\pm$0.007) & 1.27 & 1.04 \nl
2 & 0.984 ($\pm$0.009) & 1.81 & 1.12 \nl

\tableline
\multicolumn{4}{c}{}\nl
\multicolumn{4}{c}{1995 October 22}\nl
\tableline
1 & 0.985 ($\pm$0.015) & 1.29 & 1.05 \nl
2 & 0.951 ($\pm$0.016) & 2.40 & 1.40 \nl

\enddata
\end{deluxetable}

\clearpage
\begin{deluxetable}{cccc}
\tablecolumns{4}
\footnotesize
\tablecaption{Properties of the intrinsic C~IV absorption \label{tbl-3}}
\tablewidth{0pt}
\tablehead{
\colhead{Component} & \colhead{N (C~IV)} & \colhead{v$_{r}$}
& \colhead{FWHM} \\
\colhead{} & \colhead{(10$^{-14}$cm$^{-2}$)} & \colhead{(km s$^{-1}$)}
& \colhead{km s$^{-1}$)} 
}

\startdata
\multicolumn{4}{c}{1995 April 24}\nl
\tableline
1 &~8.4 $^{+0.5}_{-0.4}$  &$-$377  &130  \nl
2 &15.3 $^{+1.6}_{-1.0}$  &$-$148  &214  \nl
3 &~0.4 $^{+0.1}_{-0.1}$  &~$-$88  &~17  \nl
4 &~0.5 $^{+0.1}_{-0.1}$  &~$-$34  &~27  \nl
\tableline
\multicolumn{4}{c}{}\nl
\multicolumn{4}{c}{1995 October 22}\nl
\tableline
1 &~6.6  $^{+0.5}_{-0.4}$ &$-$377  &123  \nl
2 &15.4  $^{+4.2}_{-2.3}$ &$-$148  &198  \nl
3 &~0.4  $^{+0.1}_{-0.1}$ &~$-$88  &~23  \nl
4 &~0.5  $^{+0.1}_{-0.1}$ &~$-$29  &~35  \nl

\enddata
\end{deluxetable}

\clearpage

\figcaption[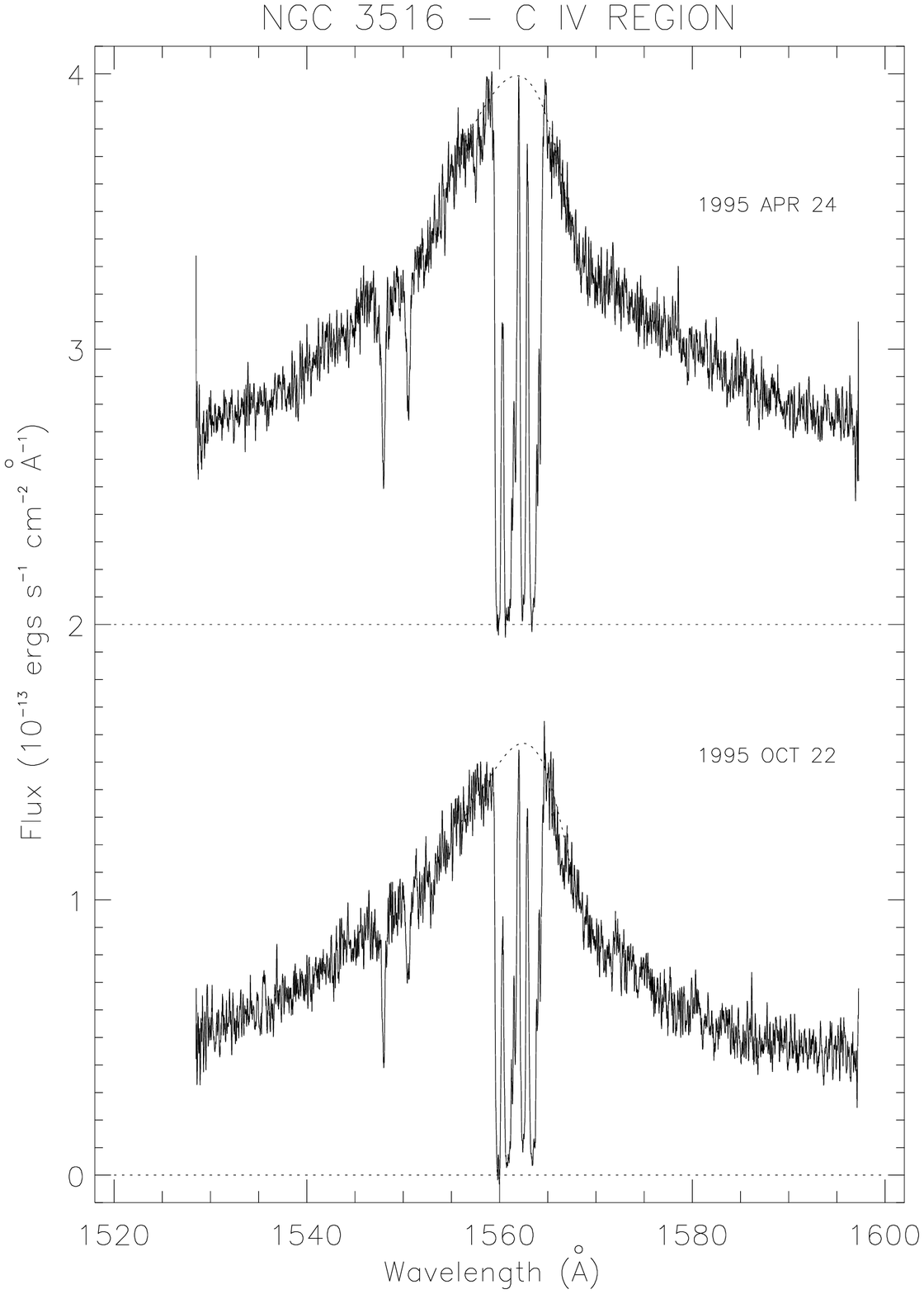]{GHRS spectra of NGC 3516 in the C IV region, obtained on 
two separate occasions. The upper spectrum is offset by 2.0 x 10$^{-13}$ erg 
s$^{-1}$ cm$^{-2}$ \AA$^{-1}$. Galactic and intrinsic C~IV absorption are 
evident in the blue wing and core of the emission profile respectively. A cubic 
spline fit to the core of the emission line is shown as a dotted line.
}\label{fig1}

\figcaption[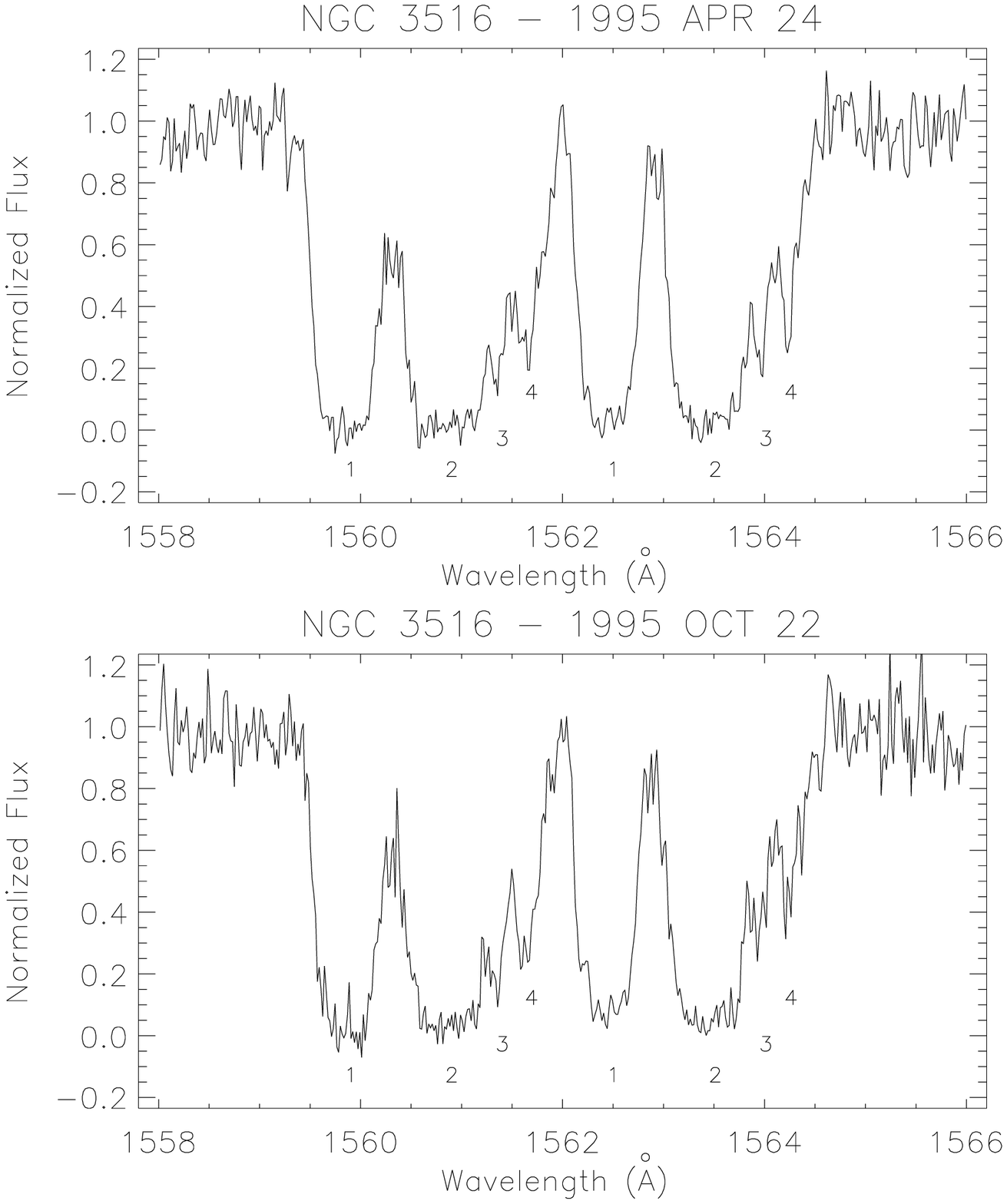]{Normalized (unsmoothed) profiles of the intrinsic C~IV 
absorption, obtained by dividing the observed spectrum by the fit to the 
emission core. Four distinct kinematic components of the C~IV 
$\lambda\lambda$1548.2, 1550.8 absorption doublet are identified.}\label{fig2}

\figcaption[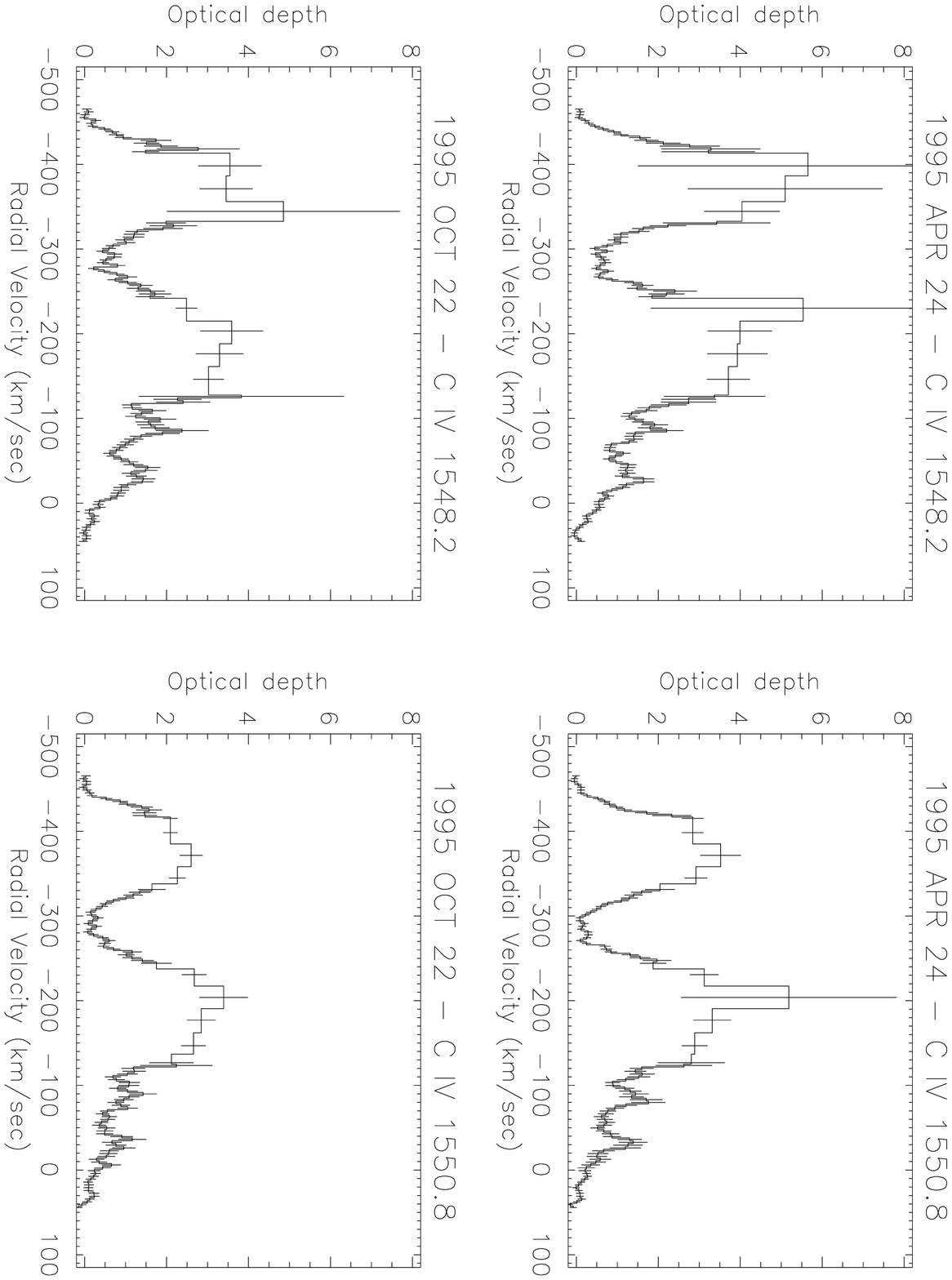]{{\it Observed} optical depths of the intrinsic C~IV 
absorption as a function of radial velocity (relative to the systemic radial 
velocity of the host galaxy.)}\label{fig3}

\plotone{fig1.eps}

\plotone{fig2.eps}

\plotone{fig3.eps}

\end{document}